\documentclass[aps,twocolumn,prb]{revtex4}
\usepackage{graphicx}

\begin{document}

\title{SUPERCHARGE   OPERATOR OF   HIDDEN  SYMMETRY IN  THE  DIRAC
EQUATION}

\author{Tamari~T.~Khachidze and Anzor~A.~Khelashvili}
\affiliation{Department of Theoretical Physics, Ivane
Javakhishvili Tbilisi State University, I.Chavchavadze ave. 3,
0128, Tbilisi, Georgia}

\email[]{khelash@ictsu.tsu.edu.ge}

\begin{abstract}

Abstract:   As is known, the so-called Dirac $K$-operator commutes
with the Dirac Hamiltonian for arbitrary central potential $V(r)$.
Therefore the spectrum is degenerate with respect to two signs of
its eigenvalues. This degeneracy may be described by some
operator, which anticommutes  with  $K$. If this operator commutes
with the Dirac Hamiltonian at the same time, then it establishes
new symmetry, which is Witten's supersymmetry. We construct the
general anticommuting with $K$  operator, which under the
requirement of this symmetry unambiguously select the Coulomb
potential. In this particular case our operator coincides with
that, introduced by Johnson and Lippmann   many years ago.

\end{abstract}

\maketitle


\ \

Supersymmetry  (SUSY)  of the Hydrogen atom is rather old and
well-studied problem. We have in mind the usual cases: SUSY in
non-relativistic quantum mechanics and inclusion of spin degrees
of freedom by Pauli method as well. In this last case the
projection of well-known Laplace-Runge-Lenz vector onto the
electron spin direction plays the role of
supercharge~\cite{Tangerman}.

   Relatively less is known about the Dirac electron,  although the so-called radial SUSY was
   demonstrated a long-time ago~\cite{Sukumar}. As for 3-dimensional case, it
   was shown~\cite{Dahl,Stahlhofen, Katsura, Khachidze}, that the supercharge
    operator is the one, introduced by Johnson and
    Lippmann in 1950 in the form of a brief abstract~\cite{Johnson}.
     As regards to the more detailed derivation, to the best our knowledge,
      is not published in scientific literature. Moreover as far as commutativity
       of the Johnson-Lippmann operator with the Dirac Hamiltonian is concerned,
        it is usually mentioned that it can be proved by  "rather tedious calculations."~\cite{Katsura}

    The main aim of our paper is a derivation of the Johnson-Lippmann operator in a simple
     and transparent manner and simultaneous demonstration of its commutativity with the
      Dirac Hamiltonian.

    Below we show that among all central potentials the Coulomb potential is
     a distinguished one. The additional symmetry takes place only for this potential.
      Then we show that the operator responsible for that symmetry reduces to the
       Johnson-Lippmann one in case of Coulomb potential.

   So, let us consider the Dirac Hamiltonian for arbitrary central potential,$V(r)$:
\begin{equation}
\label{1}
 H = \vec {\alpha}  \cdot \vec{ p} + \beta m + V(r)
\end{equation}

In this form  $V(r)$   is a fourth component of a Lorentz
4-vector. We mention this fact here because the pure Lorentz
-scalar potential is also often considered~\cite{Critchfield}.

 Let us introduce the so-called Dirac operator~\cite{Dirac},
\begin{equation}
\label{2} K = \beta \left( {\vec {\Sigma}  \cdot {\vec l} + 1}
\right) \end{equation}
where $\vec{l}$  is the angular momentum
vector, $\vec{\alpha}$ and ${\beta}$ are the usual Dirac matrices
and $ \vec { \Sigma }$ \ is the electron spin matrix

\begin{equation}
\label{3}
 \vec {\Sigma}  = \left( \begin{array}{l}
 \vec {\sigma} \ 0 \\
 0 \ \vec {\sigma}  \\
 \end{array} \right) \
 \end{equation}

It is easy to show that
\begin{equation}
\label{4}
 \left[ K , H \right] = 0
      \end{equation}
for arbitrary central potential, $V(r)$.

    Therefore the spectrum of Dirac equation is degenerate with respect to eigenvalues
    $\kappa$  of the $K$  -operator. As a rule, this is a degeneracy with regard to
     the signs of ${\kappa}$, $(\pm {\kappa})$~\cite{Sakurai}.

    We can find an operator $A$, which could  connect these two signs.
     Naturally, such an operator should anticommute with $K$,
\begin{equation}
\label{5}
\{ A, K \} = AK + K A = 0
\end{equation}

    If at the same time this operator commutes with Hamiltonian $H$,
     it'll generate the symmetry of the Dirac equation.

Therefore, we are looking for an operator $A$ with the following
properties
\begin{equation}
\label{6}
 \left[ {A,H} \right] = 0,\ \left\{ {K,A} \right\} = 0
 \end{equation}

    After that we will be able to construct supercharges as follows~\cite{Dahl,Stahlhofen, Katsura,Khachidze}
\begin{equation}
 \label{7}
 Q_1  = A,~~~~~~ \ Q_2  = i\frac{{AK}}{{\left| \kappa  \right|}}
 \end{equation}
Then it is obvious that
\begin{equation}
\label{8}\left\{ {Q_1 ,Q_2 } \right\} = 0 \end{equation}
and we
can construct Witten's  superalgebra, where $Q_1^2=Q_2^2\equiv h$
is a Witten's Hamiltonian.
      Now our goal is a construction of the $A$   operator. For this purpose
      at first we generalize one theorem~\cite{Tangerman,Biederharn},
       known from Pauli equation to the
      case of Dirac equation. For the Dirac case this theorem may be formulated as follows:

    \textbf{Theorem:}
     \textit{Suppose $\vec{V}$  be a vector with respect to the angular momentum $\vec{l}$,
     i.e.}
      $$\left[ {l_i ,V_j } \right] = i\varepsilon _{ijk} V_k$$
\textit{or, equivalently, in the vector product form one has}
 $$\vec l \times
\vec V + \vec V \times \vec l = 2i\vec V$$ \textit{Suppose also
that this vector is perpendicular to} $\vec{l}$ $$\left( {\vec l
\cdot \vec V} \right) = \left( {\vec V \cdot \vec l} \right) =
0.$$

\textit{Then $K$  anticommutes with
operator$(\vec{\Sigma}\cdot\vec{l})$ ,
   which is scalar with respect to the total $\vec{J}$  momentum, i.e.
    it commutes with $\vec{J}=\vec{l}+1/2\vec{\Sigma}$.}\\

    The proof of this theorem is almost trivial - it is sufficient to
     consider the product \ $(\vec{\Sigma} \cdot \vec{l})$  \ in this and in reversed orders and make
      use of definition of $K$. Then it follows that
\begin{equation}
\label{9}
\left\{ {K,\vec \Sigma  \cdot \vec V} \right\} = K\left(
{\vec \Sigma  \cdot \vec V} \right) + \left( {\vec \Sigma  \cdot
\vec V} \right)K = 0
\end{equation}
     It is evident that the class of operators anticommuting with $K$
     (so-called, K-odd operators) is not restricted by these operators only.
      Any operator of the form $\hat{O}(\vec{\Sigma}\cdot\vec{l})$ , where $\hat{O}$ commutes with
      $K$,   but otherwise arbitrary, also is a $K$  -
      odd.
           Let us remark for the further application, that the following
       useful relation holds in the framework of conditions of the above
       theorem
\begin{equation}
\label{10} K\left( {\vec \Sigma  \cdot \vec V} \right) =  - i\beta
\left( {\vec \Sigma  \cdot \frac{1}{2}\left[ {\vec V \times \vec l
- \vec l \times \vec V} \right]} \right)
\end{equation}
   Now one can proceed to the second stage of our problem - we wish to construct
   the K-odd operator A, which commutes with $H$. It is clear that there remains
   large freedom according to the above mentioned remark about $\hat{O}$ - operator
   - one can take $\hat{O}$  into account or ignore it.

      We have the following physically interesting vectors at hand which obey
       the requirements of our theorem. They are:\\

            $\hat {\vec {r}}$  - unit radius vector    and    $\vec{p}$    - linear momentum
            vector.\\

             Both of them are perpendicular to $\vec{l}$. Constraints of
this theorem are also satisfied by Laplace-Runge-Lenz vector
$\vec{A}=\hat {\vec{r}}-\frac{i}{2ma}[\vec{p}\times
\vec{l}-\vec{l}\times \vec{p}]$, but this vector is associated to
the Coulomb potential. Hence, we abstain from its consideration
for now. We can remark that $(\vec{\Sigma}\cdot\vec{A})$  is not
an independent structure. It is expressible by two other
structures, e.g.\
$(\vec{\Sigma}\cdot\vec{A})=(\vec{\Sigma}\cdot\hat{\vec{r}})+\frac{i}{ma}\beta
K(\vec{\Sigma}\cdot\vec{p})$\ . Therefore, we choose the following
$K$-odd terms:
  \begin{eqnarray}
  \label{11}
                  (\vec{\Sigma}\cdot\hat{\vec{r}}) {\rm  \ ~~ and~~ \ }
                  K(\vec{\Sigma\cdot\vec{p}})
                  \end{eqnarray}

  As it turns out inclusion of $K$ into the second term of (11) is necessary for obtaining our final result.
    Let's remark that both operators in (11) are diagonal matrices, while the
    Hamiltonian (1) is nondiagonal. Therefore, in commuting of (11) with nondiagonal
terms appear as well. For
     instance,
\begin{equation}
\label{12}
 \left[ \vec \Sigma  \cdot \hat {\vec {r}},H \right] =
\frac{{2i}}{r}\beta K\gamma ^5
\end{equation}
Therefore, we probe the following operator
 \begin{equation}
 \label{12}
 A = x_1 \left( \vec
\Sigma  \cdot \hat {\vec {r}} \right) + ix_2 K\left( {\vec \Sigma
\cdot \vec p} \right) + ix_3 K\gamma ^5 f\left( r \right)
\end{equation}

    Here the coefficients $x_i \left( i=1,2,3 \right)$ are chosen in such a way that $A$ is Hermitian
     operator for arbitrary real numbers
     and $f\left( r \right)$
      is an arbitrary scalar function to be determined later.

    The commutator of $A$ with $H$ is calculated straightforwardly,  the result
    is\\

$ \left[ {A,H} \right] = x_1 \frac{2i}{r}\beta K\gamma ^5 + x_2
K\left( \vec {\Sigma}  \cdot \hat {\vec {r}} \right)V'\left( r
\right)
-
\\
  - x_3 K\left( \vec {\Sigma}  \cdot \hat {\vec {r}} \right) f' \left( r \right) - ix_3 2m\beta K\gamma ^5
  f\left( r \right)
  $\\

Equating the above expression to zero, i.e. requiring
commutativity of our operator with the Dirac Hamiltonian, we find\
\begin{eqnarray}
\label{14} K\left( \vec \Sigma  \cdot \hat {\vec {r}}
\right)\left[ x_2 V'\left( r \right) - x_3 f'\left( r \right)
\right] + \\
  + 2i\beta K\gamma ^5 \left[ \frac{x_1 }{r} - mx_3 f\left( r \right) \right] =
  0 \nonumber
\end{eqnarray}

Here terms are grouped in a way that we have a diagonal matrix in
the first row, while the anti-diagonal matrix is in the second
row. Therefore, the two equations follow:
\begin{equation}
\label{15}
x_2 V'\left( r \right) = x_3 f'\left( r \right)
\end{equation}
\begin{equation}
\label{16}
 x_3 mf\left( r \right) = \frac{{x_1 }}{r}
\end{equation}

    Integration of the Eq. (15) with the requirement that  functions $f\left( r\right)$ and  $V\left( r\right)$
    tend to zero when $r \rightarrow \infty$, yields
\begin{equation}
\label{17}
x_3 V\left( r \right) = x_3 f\left( r \right)
\end{equation}
while the equation (16) gives
\begin{equation}
\label{18} f\left( r \right) = \frac{{x_1 }}{{x_2 }}\frac{1}{m r}
\end{equation}

     Substituting  Eq.  (18) into Eq   (17)  results in the following
     potential
     \begin{equation}
     \label{19}
     V\left( r \right) = \frac{{x_1 }}{{x_2 }}\frac{1}{m r}
     \end{equation}

    \textit{Hence, in the very general framework we have shown that the only central potential for
    which the Dirac Hamiltonian
     would have an additional symmetry (in the above mentioned sense) is a} \textbf{ Coulomb
     potential.}

     Meanwhile, the relative signs of coefficients $x_1$ and $x_2$ may be arbitrary. Therefore we have a  symmetry
     both for attraction and repulsion.
    If we take the Coulomb potential in the usual form
\begin{equation}
\label{20} V_c \left( r \right) =  - \frac{a}{r}
\end{equation}
where $a = Ze^2  = Z\alpha$ , it follows
\begin{equation}
\label{21}
x_2 = - \frac{1}{{ma}}x_1
\end{equation}

     In this case our symmetry operator (13) becomes
\begin{equation}
\label{22} A = x_1 \{ \left( \vec \Sigma  \cdot \hat { \vec{ r}}
\right) - \frac{i}{{ma}}K\left( {\vec \Sigma  \cdot \vec p}
\right) + \frac{i}{{mr}}K\gamma ^5  \}
\end{equation}
Number $x_1$, as an unessential common factor may be omitted.
Moreover, if we make transition to the usual Dirac $\vec \alpha $
matrices according to the relation $ \vec \Sigma  = \gamma ^5 \vec
\alpha $, then operator $A$   can be reduced to the form
\begin{equation}
\label{23}A = \gamma ^5 \{ \vec \alpha  \cdot \hat {\vec {r}} -
\frac{i}{ma}K\gamma ^5 \left( H - \beta m \right) \}
\end{equation}
 which
coincides precisely with the Johnson-Lippmann
operator~\cite{Johnson}.

   We mention here that if the potential in the Dirac Hamiltonian was a Lorentz-scalar
   (which means the change $V \rightarrow \beta V$ )  then, while $K$ still commutes with $H$,
    operator $A$ does not commute anymore with  $H$ even for Coulomb potential.
    Thus, we are convinced that in this problem of supersymmetry, the Coulomb potential
     (as a fourth component of 4-vector, i.e. minimal gauge invariant switching) takes exceptional role -
    the supercharges and Witten algebra can be constructed only for this potential.
   What the real physical picture is standing behind this?
Remark that taking into account the Eq. (10) for $ \vec V = \vec
p$, one can recast our operator for the Coulomb potential in the
following form
\begin{equation}
\label{24} A = \vec \Sigma  \cdot \left( \hat { \vec {r}} -
\frac{i}{2ma}\beta \left[ \vec p \times \vec l - \vec l \times
\vec p \right] \right) + \frac{i}{m r}K\gamma ^5
\end{equation}
One can see that in the non-relativistic limit, i.e. $\beta
\rightarrow 1$   and $ \gamma^5 \rightarrow 0 $ , our operator
reduces to
\begin{equation}
\label{25}A_{NR}  = \vec \sigma  \cdot \left( \hat {\vec {r}} -
\frac{i}{2 m a}\left[ {\vec p \times \vec l - \vec l \times \vec
p} \right]\right)
\end{equation}
 Note the Laplace-Runge-Lenz
vector in the parenthesis of Eq. (25). Therefore, relativistic
supercharge reduces to the projection of the Laplace-Runge-Lenz
vector on the electron spin direction. Precisely this operator,
Eq.(25) was used in the case of Pauli electron~\cite{Tangerman}.

     Therefore, we see that there is a deep relation between supersymmetry of the
      Dirac Hamiltonian and the symmetry related to the Laplace-Runge-Lenz vector, which
      appeared already in classical mechanics and provides the closeness of celestial orbits.

     We can conclude that the hidden symmetry associated to the Laplace-Runge-Lenz vector governs
     very wide range of physical phenomena from planetery motion to fine and hyperfine structure of
     atomic spectra. As for the Lamb shift, which is pure quantum field theory effect, its Hamiltonian,
     derived by radiative corrections to a photon propagator and photon-electron vertex function, does not
     commute with $A$  operator  and therefore spoils the above mentioned symmetry (supersymmetry). In other words,
     symmetry of the $A$  operator  controls the absence of the Lamb shift in the Dirac theory.

    In conclusion, one must also remember that the form of obtained symmetry operator is not
    unique. One can always replace $A \rightarrow \hat{O} A$ ,  where   $ [\hat{O}, H] = 0 $  and $ [\hat{O}, K] = 0 $ . One can take,
     for example, $ \hat{O}= f \left( K \right )$  - arbitrary regular
     matrix function of $K$  . Moreover, SUSY in specific and mostly exotic models of Dirac equation (such as
      2+1 dimensions~\cite{Ui}, non-minimal or anomalous magnetic moment coupling~\cite{Semenov},
      squared equation~\cite{DeJonghe, Horvathy}, etc.) are not
      excluded by our above consideration.\\

\end{document}